\documentclass[amsmath,amssymb,pra,final,showpacs,twocolumn]{revtex4-1}

\usepackage{bm,hyphenat,xspace}
\usepackage{graphicx,epsfig}

\newcommand {\mcu}{\mathcal{U}}

\begin{document}

\title {Efimov physics in bosonic atom-trimer scattering}

\author{A.~Deltuva} 
\affiliation{Centro de F\'{\i}sica Nuclear da Universidade de Lisboa, P-1649-003 Lisboa, Portugal }

\received{August 10, 2010}

\begin{abstract}
Bosonic atom-trimer scattering is studied in the unitary limit
using momentum-space equations for four-particle transition operators. 
The impact of the Efimov effect on the atom-trimer scattering
observables is explored and a number of universal relations 
is established. Positions and widths of tetramer resonances
are determined. Trimer relaxation rate constant is calculated.
\end{abstract}

 \maketitle


Few-particle systems with resonant interactions, characterized by
the two-particle scattering length $a$ being much larger than the range
of the interaction, were predicted to have a number of universal 
(interaction-independent) properties
and correlations between observables 
\cite{efimov:plb,braaten:rev,dincao:09a,hammer:07a,stecher:09a,PhysRevLett.102.133201}.
Such a behavior was recently confirmed in cold atom physics experiments
\cite{kraemer:06a,ferlaino:09a},
but can be seen qualitatively also in few-nucleon systems 
\cite{efimov:plb,braaten:rev}.
One of the best-known examples is the existence of the infinite number of
weakly bound three-boson states (Efimov trimers) in the 
unitary limit $a=\infty$ \cite{efimov:plb}.
In that limit, depending on the available energy,
an infinite number of atom-trimer channels may be present
in the four-boson system.
Due to complexity of the multichannel four-particle scattering problem
the universal properties of such a system,
i.e., the atom-trimer continuum, are not known yet;
we therefore aim to study them for the first time in the present work.
Furthermore,  the existence of a pair of four-boson states (tetramers) 
associated with each Efimov trimer was predicted \cite{hammer:07a,stecher:09a}.
However, only the two tetramers associated with the trimer
ground state are true bound states that have been studied in all the details
using standard bound state techniques 
\cite{dincao:09a,hammer:07a,stecher:09a,platter:04a}.
Higher tetramers are resonances above the atom-trimer threshold
and for this reason their properties are far less known;
they will be determined in the present work using proper scattering
calculations.

Our description of the four-boson scattering is based on the 
Alt, Grassberger and Sandhas (AGS) equations \cite{grassberger:67} 
for the transition operators;
they are equivalent to the Faddeev-Yakubovsky equations \cite{yakubovsky:67}
for the wave-function components.
Symmetrized form of AGS equations \cite{deltuva:07a} 
is appropriate for the system of four identical bosons,
\begin{subequations} \label{eq:U}
\begin{align}  
\mcu_{11}  = {}&  P_{34} (G_0  t  G_0)^{-1}  
 + P_{34}  U_1 G_0  t G_0  \mcu_{11} + U_2 G_0  t G_0  \mcu_{21} , 
\label{eq:U11} \\  
\mcu_{21}  = {}&  (1 + P_{34}) (G_0  t  G_0)^{-1}  
+ (1 + P_{34}) U_1 G_0  t  G_0  \mcu_{11} , \label{eq:U21}
\end{align}
\end{subequations}
where $\mcu_{\beta\alpha}$ are the four-particle transition operators,
$G_0$ is the four free particle Green's function,  and
$P_{34}$ is the permutation operator of  particles $3$ and $4$
that ensures correct permutation symmetry  of the system.
The dynamic input is the two-boson potential $v$ from which
the two-particle transition-matrix $t$ and
the AGS transition operators $U_{\alpha}$ are derived,
with $\alpha=1$ and 2 corresponding to the
 $1+3$ and $2+2$ subsystems, respectively. 
As explained in Ref.~\cite{deltuva:07a},
the atom-trimer scattering amplitudes are given by the
on-shell matrix elements of $\mcu_{11}$ calculated between
the Faddeev amplitudes of the corresponding initial and final states.

We solve AGS equations using momentum-space partial-wave representation
\cite{deltuva:07a}
where they are a system of coupled three-variable integral equations
that after the discretization of momentum variables
becomes a very large system of linear algebraic equations. 
In the case of the four-nucleon scattering
those equations have been successfully solved with realistic nuclear and
Coulomb interactions \cite{deltuva:07b,deltuva:08a}.
 In the present calculations we take the
numerical techniques for the treatment of four-particle permutations
and trimer bound-state poles from 
Refs.~\cite{deltuva:07a,deltuva:07b,deltuva:08a}. 
However, an important difference as compared to the four-nucleon system
is the presence of sharp resonances in the four-boson system,
manifesting themselves as poles of the AGS operators \eqref{eq:U}.
Since the convergence of the multiple scattering series in the vicinity
of the pole is very slow, in this work we solve systems of linear equations 
by the direct matrix inversion instead
of the iterative double Pad\'{e} summation method \cite{deltuva:07a}.
Since we are interested in the universal properties that
must be independent of the interaction details,
we use rank 1 separable two-boson potentials
$v = |g\rangle \lambda \langle g| $ acting in $S$-wave only
and thereby reducing Eqs.~\eqref{eq:U}
to a small system of two-variable integral equations. 
Although the  two-boson interaction is limited to $S$-wave,
i.e., $l_x=0$ in the notation of Ref.~\cite{deltuva:07a},
higher angular momentum states with $l_y, l_z \le 2$
 have to be taken into account
for the relative motion in $1+2$, $1+3$, and $2+2$ subsystems
to achieve the convergence.

Our standard potential has simple gaussian form factor, i.e., 
its momentum-dependence is
$\langle k |g\rangle = e^{-(k/\Lambda)^2}$.
The strength $\lambda$ is chosen to reproduce infinite two-boson scattering 
length. In that limit 
all observables scale with $\Lambda$; e.g., the binding energy
of the $n$-th excited trimer $b_n \sim \Lambda^2$. 
It therefore makes no sense to specify particular value of $\Lambda$ as 
well as boson mass. Instead, we will use dimensionless ratios.
As the length scale associated with the $n$-th excited trimer
we will use $l_n = \hbar / \sqrt{2\mu_1 b_n}$ 
where $\mu_1$ is the reduced atom-trimer mass.

\begin{table}[!]
\begin{ruledtabular}
\begin{tabular}{*{4}{c}} $n$ & 
 $b_{n-1}/b_{n}$ &  $\kappa_{n-1}/\kappa_n$ & $b_n/\kappa_n$
\\  \hline
1 & 548.114 & 32.734 & $1.03\times 10^{-2}$ \\
2 & 515.214 & 23.697 & $4.73\times 10^{-4}$ \\
3 & 515.036 & 22.767 & $2.09\times 10^{-5}$ \\
4 & 515.035 & 22.699 & $9.22\times 10^{-7}$ \\
5 & 515.035 & 22.695 & $4.06\times 10^{-8}$ \\
\hline
1 & 2126.36 & 11.349 & $2.14\times 10^{-3}$ \\
2 & 518.570 & 21.528 & $8.87\times 10^{-5}$ \\
3 & 515.042 & 22.655 & $3.90\times 10^{-6}$ \\
4 & 515.035 & 22.694 & $1.72\times 10^{-7}$ \\
\end{tabular}
\end{ruledtabular}
\caption{ \label{tab:b-n}
Trimer properties  obtained with one-term (top)
and two-term form factor potentials (bottom).}
\end{table}

We do not include explicit three-body force, however, many-body forces
are simulated by a different off-shell behavior of the two-body potential.
To prove that this doesn't change universal properties of the 
four-boson system we also use the potential II with two-term form factor
$\langle k |g\rangle = [1+c_2\,(k/\Lambda)^2]e^{-(k/\Lambda)^2}$;
large negative value of $c_2=-9.17$ ensures very different off-shell behavior.
In the configuration space representation the potential II has 
 several attractive and repulsive regions much like
the one of Ref.~\cite{deltuva:10b} and supports a deeply bound trimer 
that is non Efimov state in contrast to excited states and all states
of the standard potential. This can be seen in Table~\ref{tab:b-n}
which collects ratios for calculated trimer binding energies $b_n$:
all $b_{n-1}/b_n$ are quite close to the characteristic Efimov value of 
515.035, the exception being $b_0/b_1$ for the potential II that is much larger.
However, the Efimov value is reached with good accuracy
only for high excited states $n\ge 3$. This is not surprising since
the Efimov condition $R_n >> \rho$ ensuring truly universal behavior, 
where $R_n$ is the size of the $n$-th trimer and $\rho$ the range of the 
interaction,
is only well satisfied for $n$ large enough whereas for the ground states 
$R_0 < \rho$  may take place \cite{deltuva:10b,lazauskas:he}.
Simultaneously $\kappa_{n-1}/\kappa_n$ converges towards 22.694  where $\kappa_n$
is the expectation value of the $n$-th excited trimer internal kinetic energy.
The ratio $b_n/\kappa_n$ is a measure for the high-momentum components
in the trimer wave function; as Table~\ref{tab:b-n} demonstrates,
it differs significantly for the two employed potentials.

\begin{table}[!]
\begin{ruledtabular}
\begin{tabular}{*{5}{c}} $n$ & 
$\mathrm{Re}(a_n)/l_n$ & $\mathrm{Im}(a_n)/l_n$ & 
$\mathrm{Re}(r_n)/l_n$ & $\mathrm{Im}(r_n)/l_n$ 
\\  \hline
1 & 31.1 & -5.18 & 3.35 & -0.043 \\
2 & 23.1 & -1.05 & 3.23 & -0.016 \\
3 & 22.7 & -1.08 & 3.22 & -0.017 \\
4 & 22.6 & -1.09 & 3.22 & -0.017 \\
5 & 22.6 & -1.09 & 3.22 & -0.017 \\
\hline
1 & 11.7 & -0.21 & 2.93 & -0.012  \\
2 & 22.5 & -1.52 & 3.22 & -0.024  \\
3 & 22.7 & -1.09 & 3.22 & -0.017  \\
4 & 22.6 & -1.09 & 3.22 & -0.017  \\
\end{tabular}
\end{ruledtabular}
\caption{ \label{tab:a-r}
Atom-trimer scattering length and effective range 
obtained with one-term (top) and two-term form factor potentials (bottom).}
\end{table}

\begin{figure}[!]
\includegraphics[scale=0.55]{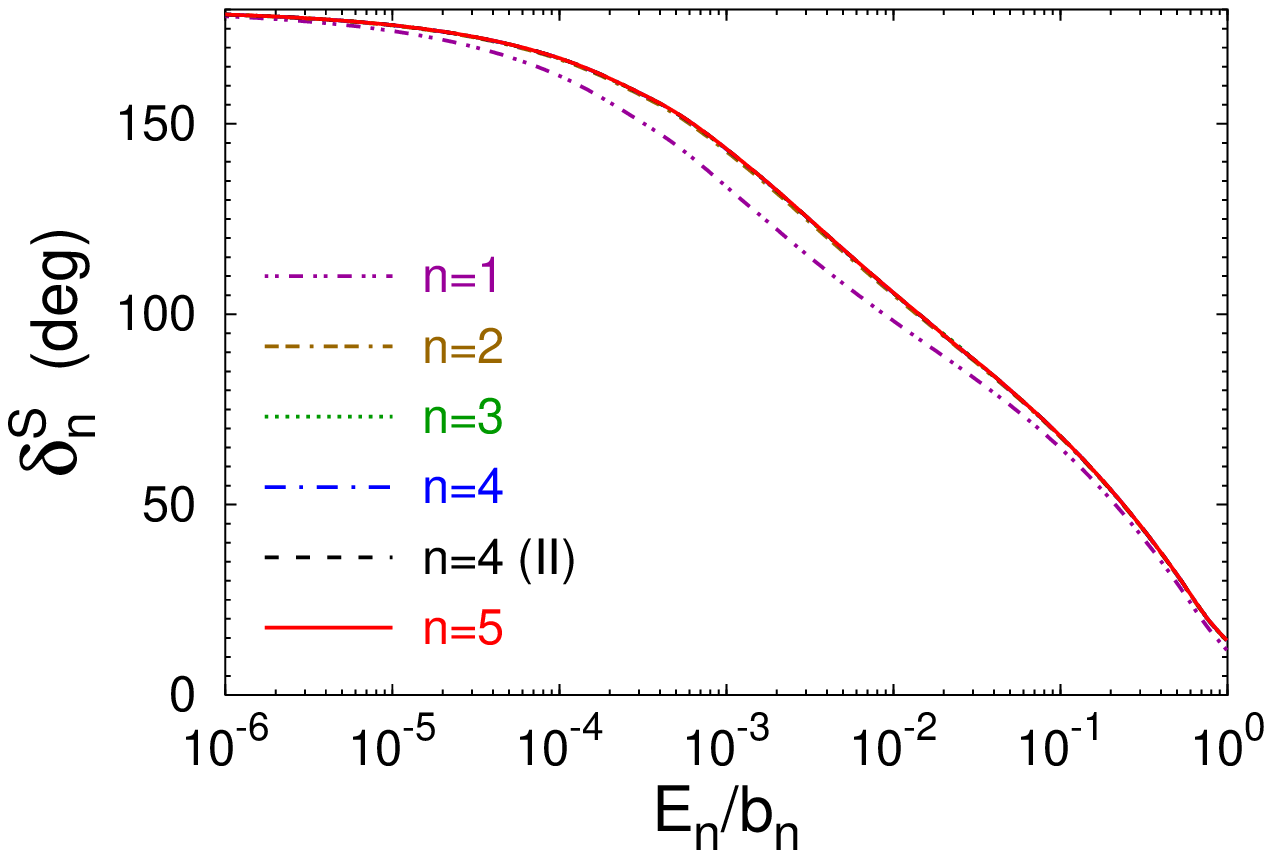}
\includegraphics[scale=0.55]{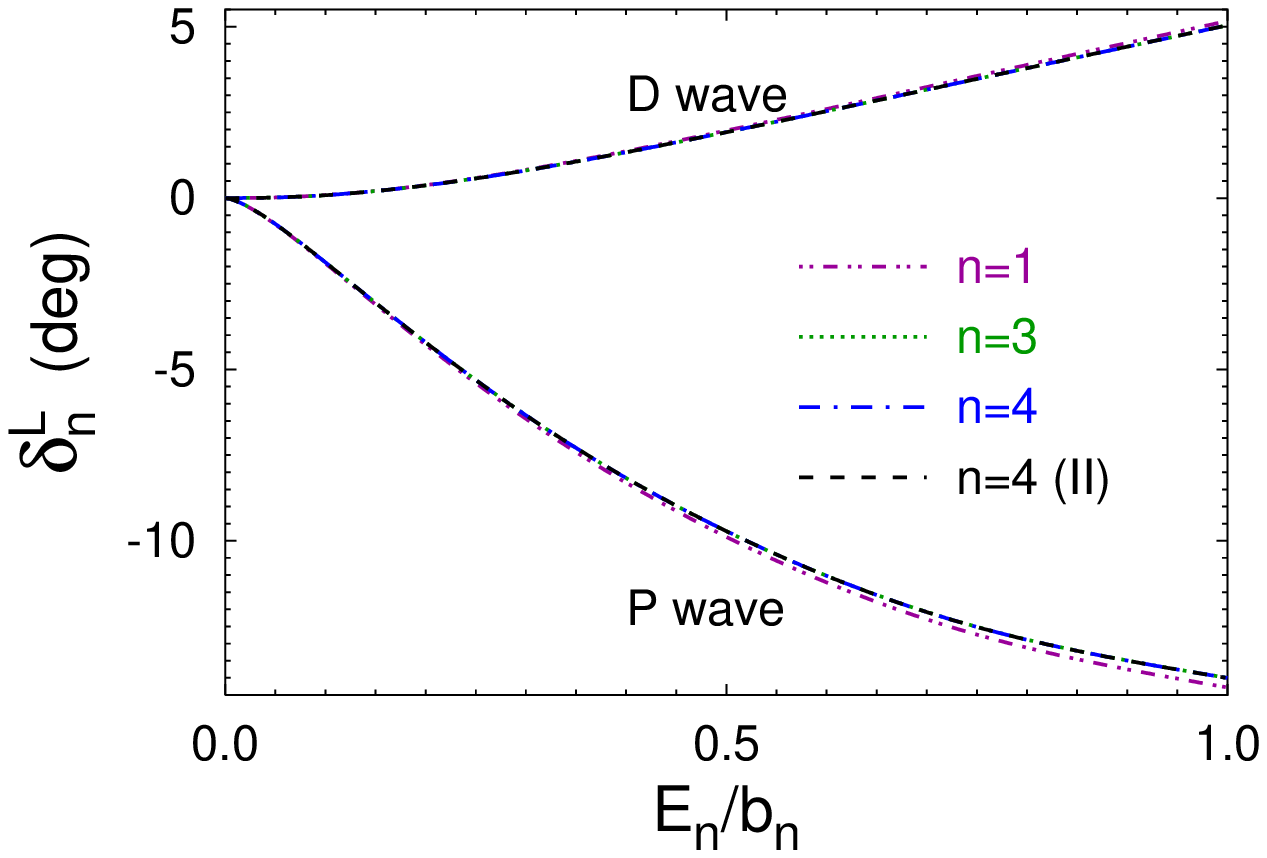}
\caption{\label{fig:phaseS} (Color online)
$S$-, $P$- and $D$-wave phase shift for the atom scattering from the $n$-th excited trimer.}
\end{figure}

\begin{figure}[!]
\includegraphics[scale=0.55]{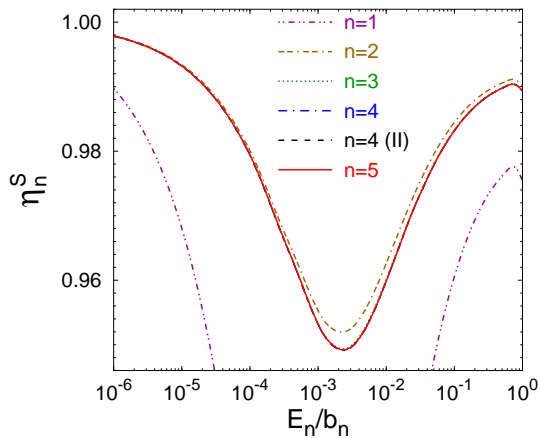} 
\caption{\label{fig:inelS} (Color online)
$S$-wave inelasticity parameter for the atom scattering from the $n$-th excited trimer.}
\end{figure}

In Table~\ref{tab:a-r} we present results for the
scattering length $a_n$ and effective range $r_n$ for the
atom scattering from the $n$-th excited trimer up to $n=5$.
For both potentials $a_n/l_n$ and $r_n/l_n$ converge towards universal values
as $n$ increases, i.e., 
\begin{subequations} \label{eq:ar-n}
\begin{align}
a_n/l_n & \approx  22.6 -1.09i, \\
r_n/l_n & \approx  3.22 -0.017i,
\end{align}
\end{subequations}
however, significant potential-dependent
deviations are observed for $n \le 2$. Including strong repulsive three-body
force that enforces the Efimov condition  $R_n >> \rho$
(but with additional numerical complications)
probably could speedup the convergence with $n$ but even in such a case
$n=0$ would be insufficient since it doesn't account for inelasticities.

It turns out that the universal limit exists for all scattering observables.
In Figs.~\ref{fig:phaseS} - \ref{fig:inelS} we show all relevant
phase shifts $\delta_n^L$
and $S$-wave inelasticity parameter $\eta_n^S$ for the atom scattering 
from the $n$-th excited trimer as functions of the relative
kinetic energy $E_n$ divided by the respective $b_n$;
the elastic $S$-matrix is parametrized as $s_n^L = \eta_n^L e^{2i\delta_n^L}$.
Again, results with $n \ge 3$ are indistinguishable and represent the
universal values, but deviations are seen for lower $n$.
Elastic scattering is determined by relative atom-trimer
$S$-, $P$-, and $D$-waves.
In fact, $S$-wave dominates at lower energies but close to  $E_n/b_n = 1$ 
the individual contributions of $S$-, $P$-, and $D$-waves
to the elastic cross section are 31, 58, an 11\%, respectively,
while the  $F$-wave with  $|\delta_n^F| < 0.4^\circ$ yields 
less than 0.1\% and therefore is negligible. 
Situation is different in the inelastic scattering where
only $S$-wave contributes significantly:
only $\eta_n^S$ clearly deviates from 1 as shown in  
Fig.~\ref{fig:inelS} while in higher
partial waves, due to very different size of trimers and the
angular momentum barrier, transitions to other trimers are
strongly suppressed and  $\eta_n^L$ are very close to 1,
e.g., $1-\eta_n^P < 10^{-6}$.

\begin{figure}[!]
\includegraphics[scale=0.62]{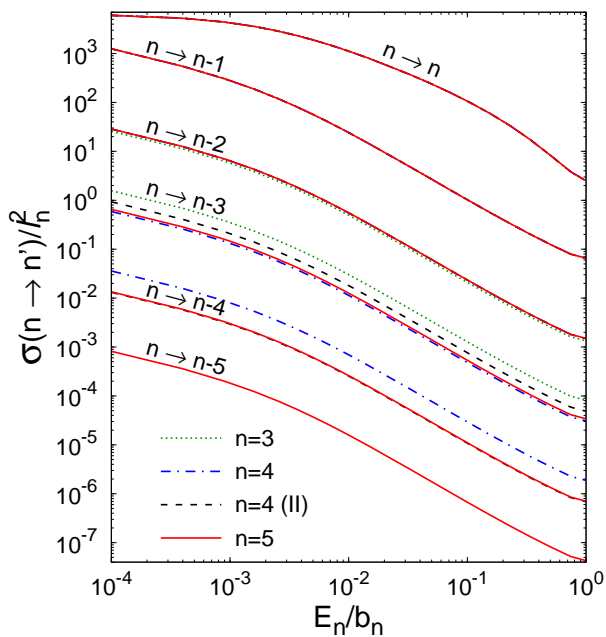} 
\caption{\label{fig:cs} (Color online)
Elastic and inelastic cross sections 
for the atom scattering from the $n$-th excited trimer.}
\end{figure}

The elastic and inelastic cross sections $\sigma(n\to n')$
for the atom scattering from the $n$-th excited trimer leading 
to the lower-lying $n'$-th trimer are presented in  Fig.~\ref{fig:cs}.  
We do not show cross sections for $n < n'$ that can be obtained
by time reversal as $\sigma(n'\to n) = (E_{n}/E_{n'})\sigma(n\to n')$.
Within the resolution of the plot
$\sigma(n\to n')/l_n^2$ for $n,n' \ge 2 $ and fixed $n-n'$ are independent
of $n$ and employed potential and thereby represent the universal values.
Furthermore, for $n > n'$ large enough the ratios
\begin{equation} \label{eq:s-nn}
\frac{\sigma(n\to n')} {\sigma(n\to n'-1)} \approx 43.7
\end{equation}
 are energy-independent. 
 These observations allow us to extrapolate
 $\sigma(n\to n')$ results to any $n$ and $n'$ that are sufficiently large.

Perhaps most interesting energy regions, namely those 
 containing $S$-wave
four-boson resonances just slightly below $E_n/b_n = 1$,
are not displayed in Fig.~\ref{fig:cs}.
There are two ($k=1,2$) four-boson resonances associated
with the $N$-th Efimov trimer. Resonances 
are poles of the AGS transition operators in the complex plane;
thus, in the vicinity of the four-boson resonance
\begin{equation} \label{eq:Upole}
\mcu_{\beta\alpha} \approx \hat{\mcu}_{\beta\alpha}^{(-1)} (E-E_r)^{-1}
+ \hat{\mcu}_{\beta\alpha}^{(0)} + \hat{\mcu}_{\beta\alpha}^{(1)}(E-E_r),
\end{equation}
where $E = E_n - b_n$ and $E_r = -B_{N,k} - i\Gamma_{N,k}/2$ with
 $-B_{N,k}$ being the $(N,k)$-th resonance position 
relative to the four-body breakup threshold and $\Gamma_{N,k}$ its width.
These parameters were determined by fitting the
on-shell matrix elements of $\mcu_{11}$ into Eq.~\eqref{eq:Upole}.
Again, for $N$ large enough, i.e., $N \ge 3$, the results with good accuracy
become independent of potential and $N$,
\begin{subequations} \label{eq:res}
\begin{align}
B_{N,1}/b_N \approx 4.6108, \;\; \quad {} & \Gamma_{N,1}/2b_N \approx 0.01484, \\
B_{N,2}/b_N \approx  1.00228,  \quad {} &
\Gamma_{N,2}/2b_N \approx  2.38\times 10^{-4}.
\end{align}
\end{subequations}
 We note that $B_{N,k}$ but not $\Gamma_{N,k}$
 have already been calculated in  Ref.~\cite{stecher:09a}.
While $B_{N,1}/b_N \approx 4.58$ of Ref.~\cite{stecher:09a} is quite close 
to our number, the $k=2$ resonance with
$B_{N,2}/b_N \approx 1.01$ was predicted in  Ref.~\cite{stecher:09a}
to be significantly further from the atom-trimer threshold than our result.
In contrast to $S$-wave, there is no four-boson 
resonances in higher angular momentum states
as can be seen in the bottom panel of Fig.~\ref{fig:phaseS}.

\begin{figure}[!]
\includegraphics[scale=0.5]{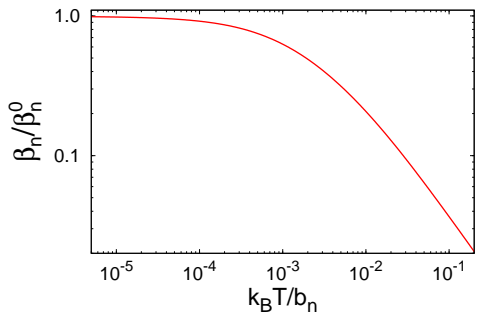} 
\caption{\label{fig:rlx} (Color online)
Temperature dependence of the trimer relaxation rate constant.}
\end{figure}

Unlike in nuclear physics, the direct measurement of the atom-trimer 
cross sections in cold atom physics experiments is not possible yet. 
Instead, one may be able to create an ultracold mixture of atoms and
excited Efimov trimers in a trap and observe the trimer relaxation, i.e.,
the inelastic collision of an atom and trimer in the $n$-th excited state
leading to the atom and trimer in the lower-lying $n'$-th state.
The kinetic energy $\Delta K \approx b_{n'}$ released in this process 
is shared between the atom and trimer with the ratio 3:1.
Thus, if $b_{n'}/4$ is larger than traping potential,
the final-state trimer escapes the trap. Assuming that this is
the dominating mechanism for the trimer loss, the time evolution of 
the density $\rho_n(t)$ of the $n$-th excited state trimers 
in the trap is given by
\begin{equation} \label{eq:rlxd}
\frac{d\rho_n(t)}{dt} = - \beta_n \rho_a(t) \rho_n(t),
\end{equation}
$\rho_a(t)$ being the atom density and $\beta_n$ the relaxation rate constant
\cite{braaten:rev}. 
The alternative way of the trimer loss, i.e.,
inelastic trimer-trimer collisions,
is suppressed if  $\rho_n(0) << \rho_a(0)$.
Under this condition $\rho_a(t) \approx \rho_a(0)$  and 
Eq.~\eqref{eq:rlxd} has a simple solution
\begin{equation} \label{eq:rlx}
\rho_n(t) = \rho_n(0) e^{-\beta_n \rho_a(0) t} .
\end{equation}
Thus, in this case the lifetime of the mixture is simply given
by $1/\beta_n \rho_a(0)$.
The relaxation rate constant $\beta_n = \sum_{n'} \beta_{n\to n'}$
has contributions 
$\beta_{n\to n'} = \langle v_n \sigma(n\to n') \rangle$
from transitions to all  trimers $n'<n$, where 
$v_n = \sqrt{2E_n/\mu_1}$ is the relative atom-trimer velocity
and $\langle \ldots \rangle$ denotes the thermal average.
Thus, the trimer relaxation rate constant is determined
by the atom-trimer inelastic cross sections calculated in the present work.
In particular, Eq.~\eqref{eq:s-nn} implies that for 
$n$ and $n'$ large enough 
$\beta_{n\to n'}/\beta_{n\to n'-1} \approx 43.7$ and therefore the relaxation
is strongly dominated by the $n \to n-1$ transition.
The zero temperature limit of the relaxation rate constant 
can be obtained using the optical theorem as
\begin{equation}  \label{eq:rlx0}
\beta_n^0 = -\frac{4\pi \hbar}{\mu_1} \, \mathrm{Im} (a_n).
\end{equation}
The results at finite temperature $T$ are given in Fig.~\ref{fig:rlx}
assuming the Boltzmann distribution for the relative atom-trimer energy;
$k_B$ is the  Boltzmann constant.
Figure~\ref{fig:rlx} indicates that the use of $T=0$ limit
is inappropriate at temperatures above $k_BT/b_n > 10^{-4}$.

In summary, we studied bosonic atom-trimer scattering in the unitary limit.
It is a complicated multichannel four-particle scattering problem
involving, in the present calculations, up to six open channels with
the Efimov trimer binding energies differing by a factor larger than 
$515^5 \approx 4 \times 10^{13}$.
Exact AGS equations were solved in momentum-space framework with some
important technical modifications compared to previous calculations of
the four-nucleon system. The results for reactions with highly 
excited trimers (at least 2nd excited state) in the initial and final channels
were found to be independent of the used potential and thereby
represent universal values
for atom-trimer scattering length, effective range, phase shifts, elastic
and inelastic cross sections and four-boson resonance parameters.
On the other hand, results for  lower trimers demonstrate that
significant quantitative deviations from the universal behavior are possible.
The comparison with the experimental data could not be performed yet,
but the obtained atom-trimer scattering results were related to
the trimer relaxation rate constant that hopefully will be measured in the 
future experiments with ultracold mixtures of atoms and Efimov trimers.

The developed technique is applicable also to dimer-dimer scattering.
Our first calculations confirm the nontrivial behavior of 
the dimer-dimer scattering length when approaching the unitary limit
\cite{dincao:09a} but can provide results also at finite energies.
Furthermore, the extension to fermionic systems in the unitary limit may
have impact not only on the cold atom but also on nuclear physics, e.g.,
by clarifying to what extent four-nucleon resonances are universal.

The author thanks R.~Lazauskas and L.~Platter for discussions
and suggestions.


\begin{thebibliography}{10}

\bibitem{efimov:plb}
V. Efimov, Phys. Lett. B {\bf 33},  563   (1970).

\bibitem{braaten:rev}
E. Braaten and H.-W. Hammer, Phys. Rep. {\bf 428},  259  (2006).

\bibitem{hammer:07a}
H.~W. Hammer and L. Platter, Eur. Phys. J. A {\bf 32},  113  (2007).

\bibitem{dincao:09a}
J.~P. D'Incao, J. von Stecher, and C.~H. Greene, Phys. Rev. Lett. {\bf 103},
  033004  (2009).

\bibitem{stecher:09a}
J. von Stecher, J.~P. D'Incao, and C.~H. Greene, Nature Phys. {\bf 5},  417
  (2009).


\bibitem{PhysRevLett.102.133201}
Y. Wang and B.~D. Esry, Phys. Rev. Lett. {\bf 102},  133201  (2009).

\bibitem{kraemer:06a}
T. Kraemer~{\it et al}, Nature {\bf 440},  315  (2006).

\bibitem{ferlaino:09a}
F. Ferlaino {\it et~al.}, Phys. Rev. Lett. {\bf 102},  140401  (2009).




\bibitem{platter:04a}
L. Platter, H.~W. Hammer, and U.-G. Mei\ss{}ner, Phys. Rev. A {\bf 70},  052101
   (2004).

\bibitem{grassberger:67}
P. Grassberger and W. Sandhas, Nucl. Phys. {\bf B2},  181  (1967); E. O. Alt,
  P. Grassberger, and W. Sandhas, JINR report No. E4-6688 (1972).

\bibitem{yakubovsky:67}
O.~A. Yakubovsky, Yad. Fiz. {\bf 5},  1312  (1967) [Sov. J. Nucl. Phys. {\bf
  5}, 937 (1967)].

\bibitem{deltuva:07a}
A. Deltuva and A.~C. Fonseca, Phys.~Rev.~C {\bf 75},  014005  (2007).

\bibitem{deltuva:07b}
A. Deltuva and A.~C. Fonseca, Phys.~Rev.~Lett. {\bf 98},  162502  (2007);
Phys.~Rev.~C {\bf 76},  021001(R)  (2007); 
Phys.~Rev.~C {\bf 81},  054002  (2010).

\bibitem{deltuva:08a}
A. Deltuva, A.~C. Fonseca, and P.~U. Sauer, Phys.~Lett.~B {\bf 660},  471
  (2008).

\bibitem{deltuva:10b}
A. Deltuva and R. Lazauskas, Phys.~Rev.~A {\bf 82},  012705  (2010).

\bibitem{lazauskas:he}
R. Lazauskas and J. Carbonell, Phys. Rev. A {\bf 73},  062717  (2006).

\end{thebibliography}

\end{document}